\documentclass[letterpaper,aps,floatfix,twocolumn]{revtex4}
\usepackage{graphicx}
\usepackage{epsfig}     
\usepackage{amsmath}
\usepackage{hyperref}
\usepackage{alltt}
\usepackage{setspace}
\newcommand{\old}[1]{{}}

\begin{document}

\title{Fractal mesh refinement, rare events and type Ia supernova}
\author{J. Glimm}
\affiliation{Stony Brook University, Stony Brook NY 11794}

\begin{abstract}

For the study of rare events, we propose a method of fractal mesh
refinement, allowing high levels of strongly focused resolution.
The method is proposed to assess the extreme events generated
by multifractal turbulent nuclear deflagration. Such events, in a white
dwarf type Ia supernova pregenitor, are assumed to lead to a deflagration
to detonation transition, which produces the observed type Ia 
supernova.

Fractal mesh refinement enables mesh refinement
regimes to the Gibson scale and beyond.
Conventional multifractal subgrid models have cost considerations
which also lead to local mesh refinement, but more importantly,
the application of these models to compressible turbulent
deflagration fronts is a topic for future research,
while subgrid models tuned to this complex physics 
lack a  multifractal search focus, and for cost reasons
do not allow the large search volumes
required to find the sought for rare event to trigger a DDT. 

Here we propose methods to resolve fine scales and locate rare possible
DDT trigger events within large volumes while addressing
multifractal issues at a feasible computational cost.
We are motivated by the goal of confirming or assessing the hypothesis of a
delayed detonation in type Ia supernova and to assessing the
delay in this event if it is found to occur.

\end{abstract}
\maketitle

\section{Introduction}
\label{sec:intro}

Type Ia super nova provide a major method to assess the distribution of mass and
relative velocities at large distances in the universe \cite{HilNie00}.
They are formed from a white dwarf which becomes unstable to nuclear
synthesis in a deflagration front while accreting mass. Due to the
very large amounts of energy released, turbulence ensues and the
deflagration fronts become unstable.  Available evidence suggests a delayed
deflagration to detonation transition (DDT). Triggering the DDT event 
may be intense fluctuations in the turbulence intensity and flame front
complexity occurring over a critical volume. Such a trigger is
a rare event. It is likely to be located in a fractal subset of
dimension $D < 4$ of space time, making the cost of a search
for the trigger prohibitively expensive in the absence of a focused search
strategy. Definitive confirmation of the DDT possibility is missing.

The phenomena of Ia supernova are summarized in
recent simulation studies \cite{ZinAlmBar17,CalKruJac12} from which
further references can be traced. Questions remain concerning the type
of pregenitor system, the nature of the burning front and its location
within the white dwarf pregenitor, the detailed mechanism leading to
a transition for deflagration to detonation (DDT) and whether this is even 
possible within common models for white dwarf pregentors.

The detailed mechanism for DDT is presumed to be pressure waves
arising from some local combustion or reaction event of extreme
intensity within a localized region. The pressure waves generated by this
event
lower the ignition temperature of the region they reach, and if the
thermodynamic conditions are very close to ignition, a wide spread
ignition and an explosion results 
according to the Zeldovich theory \cite{Lee08}. 

The trigger for DDT is presumed to be a burning front of high
turbulent intensity embedded in a larger region which is close to the
ignition temperature. The burning intensity is primarily determined
by the length of the burning front, so that the trigger can be postulated
as a local extreme of the burning front length. The burning front is unstable
to wrinkling at scales above the Gibson scale, and the instability
has been estimated to lead to a fractal burning front with a spatial
dimension $D_f \sim 2.3 > 2$. The local intensity should increase
when multifractal (multiple simultaneous length scales for wrinkling)
are considered, leading to a smaller fractal dimension and a more
intense localized combustion hot spot.

Simulation studies depend sensitively on subgrid models, which introduce 
subgrid properties of the turbulent mixing of burned and unburned stellar
material, the turbulent adjusted flame speed and the turbulent deflagration.
It is to the subgrid modeling that this short note is addressed.
A major complication to subgrid modeling is the juxtaposition
of multifractal turbulence coupled to complex physical processes.
This subgrid complication is common to a wide range of problems in
computational physics and engineering.

The multifractal physics of turbulence \cite{Fri95} is associated with
isolated or rare
events of turbulent intensity due to clustering of turbulence volatility.
Not only do regions of high turbulent intensity cluster, but there is a
compound clustering, so that the clusters themselves cluster. This compound
clustering extends to all orders, with a clustering of clusters of clusters,
etc. This clustering hierarchy of turbulent intensity
is called multifractal turbulence. Its overall
strength and importance is measured by the energy dissipation rate $\epsilon$,
which is large in the supernova context. The decrease in fractal dimension
of these compound clustering events saturates at some limiting
dimension $D_\infty > 2$, and is nearly constant beyond multiple
clustering of order 8 or so.

The synthetic reconstructions of
turbulent velocity fields or velocity gradients which follow this
multifractal statistics are called surrogate models (of multifractal
turbulence), see \cite{ZhaDinShi11} and references cited there.
The surrogate models serve as subgrid models to a hydro simulation,
and introduce the intense turbulent effects as a stochastic sub grid scale
(stochastic SGS) model.
Such models also have a computational cost, when
extended a few mesh levels beyond the feasible hydro mesh.
A larger issue is the fact that,
while well developed and validated for studies of 
single fluid incompressible turbulence,
the extension of surrogate turbulence models
to compressible turbulent deflagration does not yet exist.

Physics adapted  subgrid
modeling of turbulent compressible deflagration fronts \cite{CalKruJac12}
has been developed, but
fractal and multifractal aspects of the modeling are not included.

Combining the multifractal incompressible single fluid stochastic subgrid models
with the physically motivated compressible
turbulent deflagration front modeling ``head-on'' is
a major multi year intellectual enterprise, and is not attempted here.
Rather we proceed in the spirit of importance sampling, commonly used
in Monte Carlo studies of rare events.

We localize the
search to the regions of most importance. Due to the narrower focus, the
cost is reduced and so further levels of mesh refinement are feasible, 
probably to the Gibson scale and smaller. Of equal importance with the
added resolution is the localization of the search within a promising
fractal set, thereby circumventing the impossible requirement for large
search volumes at high resolution.

We explore an aggressive mesh refinement program
which allows the hydro solvers to achieve both objectives simultaneously, 
the high resolution and the search location restricted within a fractal set,
with a control over the total cost of the simulation.
The refined resolution of events of high turbulent intensity,
those of high order multifractal clustering, are
identified in a hierarchical manner. 

As with importance sampling, control of computational cost is
a central issue. Our strategy is a sub case of
conventional automatic mesh refinement
(AMR), which is to refine wherever solution accuracy indicates a need.
We call the proposed method fractal mesh refinement (FMR). 

Depending on parameters chosen for FMR, we can achieve an order of magnitude
of improved resolution,
to the Gibson scale, and finer. The Gibson scale, about $10^4$ cm, is
the smallest scale at which the turbulence causes the flame front to 
wrinkle. However, the flame fronts may well contain closely spaced but
smooth and nearly parallel regions of alternating burned and unburned 
material. For example, if a flame front is wound around a vortex spiral,
the front may be relatively smooth (below the Gibson limit), but with
narrow spacing between neighboring fronts. For this scenario, the
separation is limited by the flame width, suggesting that
turbulence and combustion phenomena below
the Gibson scale could be important to DDT.

The search for intense turbulent fluctuation is sensitive to the
volume over which the search occurs, and thus to the time of the search,
as the laminar spherically symmetric flame from moves outward in the
white dwarf. Thus the methods proposed have the potential not only to 
assess the delayed DDT, but to yield an estimate on the delay itself. It is
evident that application of this methodology would not
yield a consistent and deterministic (high probability)
DDT event when applied at the very beginnings of the deflagration,
due to the small volumes involved. Thus some level of delay appears
to be required.
The FMR method is applicable to other type Ia supernova
scenarios. It can be used as a verification and calibration tool
for conventional subgrid models in regions of intense turbulence. 
More broadly, FMR is
applicable to problems which require assessing extreme events,
assuming only that a knowledge of the nature of the event as observed at
each length scale is known.

\section{FMR parameters and costs}
\label{sec:fmr}

AMR is a numerical algorithm and software tool that allows refinement of
selected space time mesh cells in a regular rectangularly gridded 
domain \cite{CalKruJac12}. Normally the refinement is by factors of
2.  The power of the method is its application to successive levels of
refinement, so that refined cells are further refined, as needed.
There is a restriction that adjacent cells should differ at most by a single
level of refinement. This tapering of the refinement levels introduces
a major cost factor into our estimates. If an isolated cell is refined
once, there is no additional refinements needed, but if one of its sub cells
is refined again, any neighbor cell must be at least singly
refined, if not refined already. For the single isolated double refined cell,
there are $2^4 = 16$ singly refined cells (including the parent of the
doubly refined cell) to be refined. 

We start our cost estimates with a base simulation of cost $b$.
As an FMR parameter, we choose a fraction $f$ of the cells at each stage to
be refined. Applying this fraction to the base simulation, we label a fraction
$f$ of the cells as refined (at level zero), and with a cost still $b$, so 
that the cost multiple is 1. At the first level of refinement, a fraction
$f$ of the zero level refined cells are refined once. The number of cells
is $f^2$ and the cost is proportional to 16, the number of space time sub cells
produced by the refinement. We let $c_0$ be this proportionality factor and
$\epsilon = 16 f $, so that the level one
refinement cost is $b c_0 f \epsilon$. At level one, there is no secondary
refinement of cells to satisfy the adjacency requirement.

At level 2, $f^3$ cells are refined. The cost is $b f (c_0 \epsilon)^2$.
At level $n$, the cost is $bf (c_0 \epsilon)^n$. $c_0$ is not small, so that
the series is divergent, and only a finite number of levels of refinement are
allowed. But the factor $\epsilon$ is normally chosen to be small, so that
the cost per refinement level is substantially reduced below that of AMR.
The result is a significant increase in the number of refinement levels allowed.

The cost of FMR relative to AMR depends on the fraction of cells to be
refined in AMR at each level of refinement. As this fraction 
is problem dependent,
no universal estimate is possible. However, with the ability to choose
$f$ as small, for example $f = .01$ or $f = .001$, and with the possibility
of applying this construction within the normal AMR sequence as well as
as an extension of it,it is likely that
FMR will allow a significant increase in the number of allowed levels of
refinement. Beyond that, the refined cells, once resolved, are
far more likely to be located in a designated fractal subspace that can
serve as a trigger for DDT.

\section{Multifractal FMR and turbulent deflagration}
\label{sec:multif}

A selection criteria based on turbulent intensity alone yields a fractal
construction. Turbulence is a fractal set with dimension  $D < 3$ and the
fractal construction will yield refined subgrid cells within it. Multifractal
constructions, and compound clustering of intensities, clusters of clusters,
occur on smaller fractal sets, which mix turbulent intensities
with disparate length scales. The construction of a multifractal algorithm
is a simple modification of Sec.~\ref{sec:fmr}. We only need to change the
definition of the filter function $f$ from intensities of turbulence to
intensities of multifractal turbulence. This is conveniently done through
structure functions of order $n$. We then define $f$ to
selects extreme values of this structure function. 

The $n^{th}$ structure function $S_n$ is a measure of high order statistical
fluctuations in the velocity field $u$, reflecting compound clustering
or intermittancy of turbulent intensity. This clustering is concentrated
on a fractal set of dimension $\zeta_n$ according to the formula
\begin{equation}
\label{eq:zeta_n}
S_n = \langle (u(x+r) - u(x))^n \rangle = |r|^{\zeta_n} \ .
\end{equation}

Both the fractal and the multifractal definitions
are turbulence centric and are better replaced with 
a combustion centric analysis.
The heat release from a deflagration front is primarily proportional to
the length of the flame front. Above the Gibson scale, the flame front is
wrinkled and is itself a fractal, thus having a fractal dimension 
$D_f > 2$. By a modification of the criteria used to define the
filter function $f$, we ensure that the mesh refined regions lie on
this fractal set, with its enhanced rate of heat release. The wrinkled
flame front itself may have a multifractal structure, so that compound
short wave length flame front convolutions and longer wave length ones 
may be superimposed, and with the contributions from multiple
wave lengths of fluctuations in the compound clustering of multifractal
fluctuations. The final proposal is to choose $f$ to be located in the
fractal set defined by a high order of multifractal wrinkling.

Fractal (but not multifractal) analysis of flame
fronts is examined both experimentally and in simulation \cite{BatTroPic17}.
Fractal dimensions for the flame front in the range 2.18 to 2.35 are
found. A more turbulent frame front regime of a broken,
not wrinkled flame front is cited, but not further explored.

The multifractal nature of the deflagration front and the multifractal
nature of the turbulence are related, as the turbulent fluctuations drive
the convolutions of the deflagration front.

Which of these fractal sets (or some new one to be delineated) will actually
contain a trigger for DDT remains to be explored. Such a test will
contribute to assessing
whether the conventional view of DDT is actually correct.

\section{Conclusions}
\label{sec:conc}

We propose a method of fractal mesh refinement, (FMR),
allowing high levels of
mesh resolution of extreme events. We have proposed events that could lead to
a trigger for the initiatation of DDT in a type Ia pregenitor.
We propose its use to assess a hypothesized scenario for DDT
in a type Ia supernova as well as other type Ia supernova
scenarios. Calibration of conventional
subgrid models in regions of intense turbulence is another application.
FMR is applicable more generally to computational physics and applied
science problems
which require assessing extreme events.

\section{Acknowledgements}

This work is supported in part by
Leland Stanford Junior University (subaward with DOE as prime sponsor).

\bibliographystyle{plain}

\begin{thebibliography}{}

\bibitem{BatTroPic17}
F.~Battista, G.~Troiani, and F.~Picano.
\newblock Fractal scaling of turbulent premixed flame fronts: application to
  les.
\newblock {\em arXive}, 2017.

\bibitem{CalKruJac12}
A.~Calder, B.~Krueger, A.~Jackson, D.~Townsley, E.~Brown, and F.~Times.
\newblock On simulating type {Ia} supernovae.
\newblock {\em aeXive}, 2012.

\bibitem{Fri95}
U.~Frisch.
\newblock {\em Turbulence: The Legacy of {A}. {N}. {K}olmogorov}.
\newblock Cambridge Univeristy Press, Cambridge, 1996.

\bibitem{HilNie00}
W.~Hillebrand and J.~Niemeyer.
\newblock Type {Ia} supernova explosion models.
\newblock {\em Ann. Rev. Astro. and Astrophysics}, 38:191--230, 2000.

\bibitem{Lee08}
J.~Lee.
\newblock {\em The Detonation Phenomena}.
\newblock Campridge University Press, 2008.

\bibitem{ZhaDinShi11}
Z.-X. Zhang, K.-Q. Ding, Y.-P. She, and Z.-S. She.
\newblock Three-dimensional synthetic turblence constructed by spatially
  randomized interpolation.
\newblock {\em Phys. Rev. E}, 84, 2011.

\bibitem{ZinAlmBar17}
M.~Zingale, A.~S. Almgren, M.~G.~Barrios Sazo, V.~E. Beckner, J.~B. Bell,
  B.~Friesen, A.~M. Jacobs, M.P. Katz, C.~M. Malone, A.~J. Nonaka, D.~E.
  Willcox, and W.~Zhang.
\newblock Meeting the challenge of modeling astrophysical thermonuclear
  explosions: Castro, maestro and the amrex astrophysics suite.
\newblock {\em ArXive}, 2017.

\end{thebibliography}

\end{document}